# A Tracking prior to Localization workflow for Ultrasound Localization Microscopy

Alexis Leconte, Jonathan Porée, Brice Rauby, Alice Wu, Nin Ghigo, Paul Xing, Stephen Lee, Chloé Bourquin, Gerardo Ramos-Palacios, Abbas F. Sadikot and Jean Provost

***Abstract*— Ultrasound Localization Microscopy (ULM) has proven effective in resolving microvascular structures and local mean velocities at sub-diffraction-limited scales, offering high-resolution imaging capabilities. Dynamic ULM (DULM) enables the creation of angiography or velocity movies throughout cardiac cycles. Currently, these techniques rely on a Localization-and-Tracking (LAT) workflow consisting in detecting microbubbles (MB) in the frames before pairing them to generate tracks. While conventional LAT methods perform well at low concentrations, they suffer from longer acquisition times and degraded localization and tracking accuracy at higher concentrations, leading to biased angiogram reconstruction and velocity estimation. In this study, we propose a novel approach to address these challenges by reversing the current workflow. The proposed method, Tracking-and-Localization (TAL), relies on first tracking the MB and then performing localization. Through comprehensive benchmarking using both *in silico* and *in vivo* experiments and employing various metrics to quantify ULM angiography and velocity maps, we demonstrate that the TAL method consistently outperforms the reference LAT workflow. Moreover, when applied to DULM, TAL successfully extracts velocity variations along the cardiac cycle with improved repeatability. The findings of this work highlight the effectiveness of the TAL approach in overcoming the limitations of conventional LAT methods, providing enhanced ULM angiography and velocity imaging.**

***Index Terms*— Contrast Ultrasound, Ultrasound Localization Microscopy (ULM), super-resolution tracking.**

## I. INTRODUCTION

Ultrasound Localization Microscopy (ULM) enables the high-resolution imaging of blood vessels by localizing and tracking a large number of individual microbubbles (MB) injected as a contrast agent in 2D [1], [2], [3] and more recently in 3D [4], [5], [6], [7]. ULM studies can be conducted with ultrafast scanners or with pre-clinical and clinical scanners [8]. ULM can provide maps of the angioarchitecture along with mean flow velocities in a wide range of vessel diameters, overcoming the ultrasound diffraction limit [9], [10], [11]. Recently, Dynamic Ultrasound Localization Microscopy (DULM) has been introduced to enable the generation of super-resolved cine-loops of the vasculature using retrospective gating [12]. DULM can be used, e.g., to generate videos of the pulsatile blood flow in the brain [12], [13], [14] or to map the micro angioarchitecture of the beating heart [15]. Additionally, functional ULM in the rodent brain [16] was also introduced based on the generation of super-resolved cine-loops of the vasculature obtained using retrospective gating based on the stimulation timing. Since ULM requires the localization of single MB both in space and time, it needs longer acquisitions, which could become prohibitive for clinical translation [17], [18], [19], [20]. While increasing the concentration can be used to shorten acquisition times, high concentrations of MB in ULM lead to compromised localization precision and accuracy in both angioarchitecture maps and blood flow measurements [10], [18], [19], [21].

Standard methods in ULM [2], [16], [22], [23], [24] and DULM [12], [14] are executed post-acquisition and are typically based on a localize-and-track (LAT) workflow: detect and localize MB in a frame-per-frame fashion and then use pairing algorithms to track them in the temporal dimension. While this approach works well at low concentrations, it rapidly breaks down when the number of MB per frame increases [25], [26], [27]. Moreover, ULM studies often assume that precise MB localization and high frame rate compared to the velocities of the MB enables accurate velocity estimations [2], [9], [11]. However, there is a lack of studies confirming the velocities obtained *in vivo,* while many studies have reported issues such as the absence of a ground truth [28] or the impact of missing positions between frames, false detections and incorrect pairing can significantly bias the velocity estimation [11], [17], [29].

Kalman filtering tracking [21], [30], [31] has also been proposed to inject physically-consistent criteria (e.g. velocity continuity) to improve the selection and accuracy of

This work was supported in part by the Institute for Data Valorization (IVADO), in part by the Canada Foundation for Innovation under Grant 38095, in part by the Canadian Institutes of Health Research (CIHR) under Grant 452530, in part by the New Frontiers in Research Fund under Grant NFRFE-2018-01312 and in part by the Natural Sciences and Engineering Research Council of Canada (NSERC) under Grant RGPIN-2019-04982. The work of A. Leconte (alexis.leconte@polymtl.ca) was supported in part by the TransMedTech Institute (TMT) and in part by the Quebec Bio-Imaging Network (RBIQ-QBIN). The work of J. Porée (jonathan.poree@polymtl.ca) was supported in part by TMT. The work of B. Rauby (brice.rauby@polymtl.ca) was supported in part by TMT, in part by IVADO and in part by the Fonds de recherche du Québec — Nature et technologies (FRQNT). The work of A. Wu (alice-3.wu@polymtl.ca) was supported in part by TMT, in part by FRQNT and in part by RBIQ-QBIN. The work of P.Xing (paul.xing@polymtl.ca) was supported in part by TMT and in part by NSERC. The work of S. Lee (stephen.lee@polymtl.ca) was supported in part by TMT and in part by the Vanier Banting postdoctoral fellowship (NSERC). The work of C. Bourquin (chloe.bourquin@polymtl.ca) was supported in part by TMT, in part by IVADO, in part by FRQNT, in part by RBIQ-QBIN and in part by the Canada First Research Excellence Fund (Apogée/CFREF). The work of G.Ramos (gerardo.ramos@mail.mcgill.ca) was supported in part by CONAHCYT and in part by Insightec This research was enabled in part by support provided by Calcul Québec (calculquebec.ca) and the Digital Research Alliance of Canada (alliancecan.ca). (Corresponding author: Jean Provost: jean.provost@polymtl.ca).

trajectories, but remains an LAT approach since a distance linking algorithm is used as input.

Different approaches have been introduced either to handle the overlapping MB signals but remained limited to the imaging of the angioarchitecture [32], [33], [34], [35], [36] or to separate MB into sub-populations of different velocity range [37] to reduce overlapping MB events. However, the MB separation increased the processing time and was associated with a 'MB doubling' effect that could bias the number of localized MB as well as the velocity [37].

Deep learning architectures have been proposed to replace the localization step [26], [27], [38], [39], [40], [41], but most of them do not provide blood flow estimation. Recently, a deep learning method (Deep-SMV) has been proposed for velocity recovery, with convincing results in this direction [29].

We propose instead to detect and segment the MB trajectories as 1-D objects in a 3D spatiotemporal volume before localizing them in a Track-and-Localize (TAL) workflow. The underlying principle is to leverage the continuity between frames, enabled by the high frame rate relative to the velocity of MB, to directly segment MB trajectories without the need for spatial and temporal separation through localization and tracking. Our approach involves a spatiotemporal Hessian analysis and centerline segmentation. This methodology bears resemblance to previous techniques that leverage signal continuity across consecutive frames. For instance, in [42], a method wherein the slope of the 'trail' left by MB displacement is enhanced to estimate velocity without localization. In [16], [22], [43], a Hessian-based vesselness filter is applied to the spatiotemporal volume (2D+time) to filter signals likely to represent MB trajectories, followed by a conventional LAT workflow. However, our approach sets itself apart by concurrently detecting and pairing MB, utilizing all available spatiotemporal information. It is important to note that our proposed approach employs a Hessian-based vesselness function, not to detect vessels themselves but rather spatiotemporal tubular structures created by the motion of the point-spread functions in spacetime, which is distinct from the standard utilization of such functions that enhance vessel structures in a 2D or a 3D space volume during post-processing [32], [44], [45], [46]. This distinction is crucial as more a priori information is known about such trajectories: their diameter corresponds to the size of the PSF, branching is not expected, and their orientation is dictated by the MB velocity.

Herein, we demonstrate that TAL matches at low concentration and overperforms at high concentration the standard LAT approaches. Specifically, we build the theoretical framework enabling the segmentation of trajectories in three-dimensional space-time and validate its performance *in silico* and *in vivo* by benchmarking it using publicly available codes and datasets [47]. We then show that, in the context of DULM, TAL outperforms the standard LAT approaches, especially at high concentration in a rat and transcranial mouse brain model. Overall, the proposed TAL method leads to systematically improved performance in terms of angioarchitecture and velocity maps and provides better and more robust cine-loops in the context of dynamic imaging. We also provide the code needed to reproduce the TAL workflow at https://github.com/provostultrasoundlab/TrackingAndLocalizationULM.

## II. Methods

In the following, we first describe our proposed TAL workflow based on a modified Hessian matrix and a thinning algorithm to segment trajectories in space-time. Then we describe the standard ULM pipeline based on the LAT workflow, which is used as a gold-standard in this study. We also describe the DULM acquisition and processing pipeline. We end this section by introducing a series of metrics to quantify the performances of the proposed methodology.

### A. Spatiotemporal Tracking of MB Trajectories

After In-phase and Quadrature complex (IQ) image formation (i.e., delay-and-sum beamforming) [48], [49], tissue was removed using a Singular Value Decomposition (SVD) clutter filter to recover signals from MB [50], [51]. The proposed tracking method then takes for input the envelope of the post-SVD IQ data, i.e., $I(\boldsymbol{x})$. Inspired by [52], where the objective is to enhance tubular structures in a spatial volume, we adapted a spatiotemporal filter to enhance the trajectories of the Point Spread Function (PSF) of the MB, which possesses a tubular geometry in space-time. A modified Hessian matrix is calculated for each voxel of the spatiotemporal volume $I(\boldsymbol{x})$:

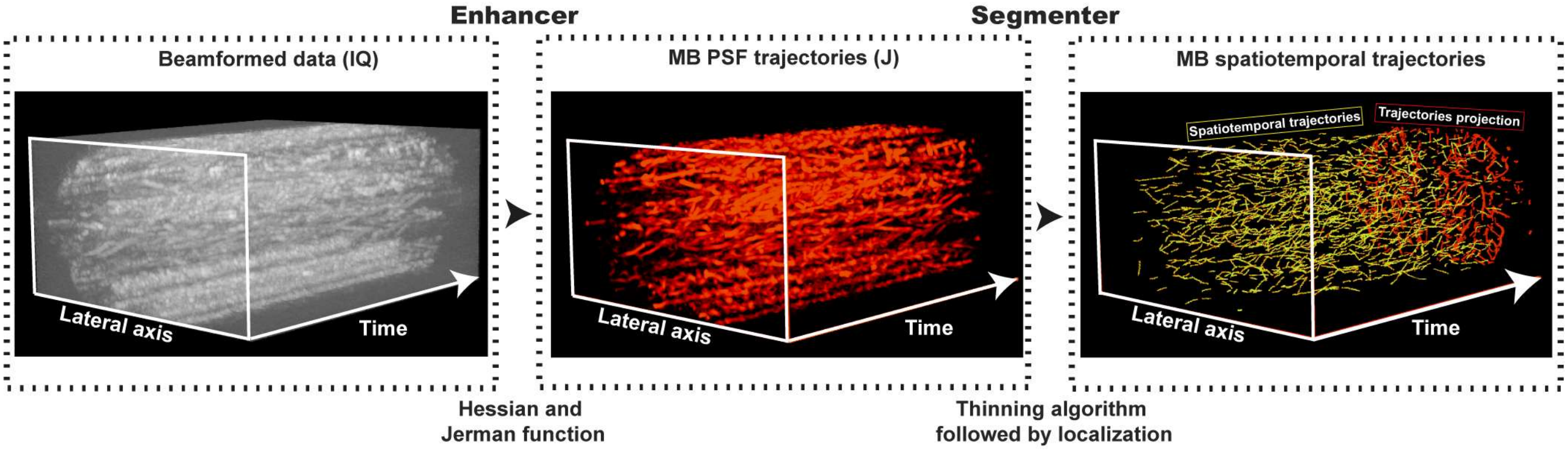

*Figure 1* **Localization and Tracking workflow (TAL)**. TAL takes a spatiotemporal (space × time) input of beamformed data, where the Jerman function is calculated (J) to extract the spatiotemporal MB PSF trajectories, before applying a thinning algorithm supported by localization algorithm to segment and super-resolve the trajectories of the moving MB.

$$H\{I\}(\boldsymbol{x}) = \lambda^2 I(\boldsymbol{x}) * \begin{bmatrix} \varepsilon^2 \frac{d^2.}{dz^2} & \varepsilon^2 \frac{d^2.}{dzdx} & \varepsilon \frac{d^2.}{dzdt} \\ \varepsilon^2 \frac{d^2.}{dxdz} & \varepsilon^2 \frac{d^2.}{dx^2} & \varepsilon \frac{d^2.}{dxdt} \\ \varepsilon \frac{d^2.}{dtdz} & \varepsilon \frac{d^2.}{dtdx} & \frac{d^2.}{dt^2} \end{bmatrix} G(\boldsymbol{x}, \lambda) \quad (1)$$

Where $\mathbf{x} = [\mathrm{z}, \mathrm{x}, \mathrm{t}]^{\mathrm{T}}$ are the spatiotemporal coordinates of the voxels in the volume, and $G(\boldsymbol{x}, \lambda) = \left(\frac{1}{\sqrt{2\pi\lambda^2}}\right)^3 exp\left(-\frac{x^T x}{2\lambda^2}\right)$ is a Gaussian kernel, where the standard deviation is defined as the width of the tubular objects, which corresponds, in our case, to the size of the PSF. While this standard deviation can be adjusted for each pixel of the image, we uniformly set it to the wavelength λ across the entire image in this work. $\varepsilon>1$ is a scalar parameter that we introduced to reject potential structures that are orthogonal to the temporal dimension, which could be enhanced because of the proximity between MB in the same frame. While $\varepsilon$ can, in principle, limit the measured velocities if chosen to be too large, detected tracks were not sensitive to its value. We show below that setting $\varepsilon$ to any value between 1.2 and 2 $\lambda$/frame is sufficient to regularize the tracking operation (see Fig.4). After diagonalization of each Hessian matrix, the Jerman function [52], developed to enhance the tubular geometry structures, is applied on the three eigenvalues $(\nu_1 \leq \nu_2 \leq \nu_3)$ :

$$J(\boldsymbol{x}) = \begin{cases} 0 & \nu_2 \leq 0 \; or \; \nu_\rho \leq 0 \\ 1 & \nu_2 \geq \nu_\rho/2 > 0 \\ \nu_2^2(\nu_\rho - \nu_2)\left(\frac{3}{\nu_2 + \nu_\rho}\right)^3 & otherwise \end{cases} \quad (2)$$

where $\nu_\rho$ is a regularized parametric version of $\nu_3$, via the $\tau \in [0,1]$ parameter, to adapt the sensitivity of the function to the non-homogeneous MB PSF trajectory's intensity:

$$\nu_\rho = \begin{cases} \nu_3 & \nu_3 > \tau \max\limits_{(z,x,t)} \nu_3(z,x,t) \\ \tau \max\limits_{(z,x,t)} \nu_3(z,x,t) & 0 < \nu_3 \leq \tau \max\limits_{(z,x,t)} \nu_3(z,x,t) \\ 0 & otherwise \end{cases} \quad (3)$$

After binarization, a thinning algorithm is used to segment the MB PSF trajectories centerline at a pixel resolution [53]. $N_{traj}$ tracks are obtained with a pixel precision $\{\boldsymbol{r}_1^i, \dots, \boldsymbol{r}_{l^i}^i\}_{i \in [\![1, N_{traj}]\!]}$, where $\boldsymbol{r} = (z, x, t)$ is the position in the space-time of a MB and $l^i$ is the length of the $i^{th}$ track. We kept only the $\tilde{N}_{traj}$ tracks longer than $N_f$ frames to increase confidence in the trajectories obtained. $\tau$, $\epsilon$ and $N_f$ are the only dataset dependent parameters. Then, from each position in each track, a localization algorithm is applied to obtain the subwavelength positions of the MB $\{\tilde{\boldsymbol{r}}_1^i, \dots, \tilde{\boldsymbol{r}}_{l^i}^i\}_{i \in [\![1, \tilde{N}_{traj}]\!]}$ (see Fig.1).

Many approaches can be used to localize the MB; here, we used the radial symmetry algorithm [47]. Finally, a function $S$ is determined for each track by applying a cubic smoothing spline as follows (Matlab function *csaps*):

$$S = \underset{S}{argmin} \; p \sum_{j=1}^{l} \|r_j - S(t_j)\|^2 + (1-p) \int \left\|\frac{d^2 S}{dt^2}(t)\right\|^2 dt \quad (4)$$

From the Plancherel theorem, we can extract a link between the $p$ parameter and a cut-off frequency $f_c$: $p = \frac{f_c^4}{1+f_c^4}$ as in [54], which will enforce the spatiotemporal smoothness of the trajectory. The representative function $S_p$ of each trajectory can be interpolated (Matlab function *fnval*) and differentiated (Matlab function *fnder*) to obtain dense trajectories and their respective velocities. The choice of the different parameters for the TAL workflow are presented in Table I.

### B. Standard LAT & LATs ULM and LATs DULM

Standard LAT and LAT + cubic smoothing spline (LATs) ULM were implemented using the publicly available PALA toolbox [47]. Briefly, after IQ image formation, individual MB were detected as local maxima in the B-mode images and localized using radial symmetry. In the PALA pipeline, a detection is rejected if there is another MB detection within a 3-λ diameter. The Hungarian method (Kuhn-Munkres algorithm) was used for tracking [55]. Different parameters need to be set to perform the tracking: the number of MB per frame that we consider (Nb), the Maximum Linking Distance to pair MB (MLD), the Maximum Frame Gap allowed for pairing (MFG) and the minimum length of a track ($N_f$) (see Table I). Subsequently, smoothing following tracking was performed in one of two ways. In standard LAT (corresponding to the publicly available PALA codes), the tracks were smoothed (Matlab function smooth) and linearly interpolated. Velocities were estimated using finite differences. In LATs, that we introduced, a function S is determined for each track by applying a cubic smoothing spline, to match the type of smoothing used in the proposed TAL ULM approach. Tracks were accumulated on a finer grid corresponding to approximately one tenth of a wavelength. LATs DULM was implemented following the methods of [12]. Briefly, retrospective gating on the ECG allows for the localization of MB in space and time within the cardiac cycle. LATs DULM tracking was performed in the same manner as for static LATs ULM, albeit detecting on the correlation map (correlation between the PSF and the post-SVD IQ data) and performing spline interpolation with adjusted parameters (Table I).

### C. In Silico localization metrics

Metrics introduced in [47] and available in the PALA toolbox were used to quantify the quality of localization and detection knowing the ground truth: 1) Localization error defined as the Root Mean Square Error (RMSE) between the i[th] simulated $(z_i^0, x_i^0)$ and detected $(z_i, x_i)$ MB positions. 2) Jaccard index: $JAC = \frac{TP}{FN+TP+FP}$, that measures the detection rate. 3) Detection precision: $Pr = \frac{TP}{FP+T}$. 4) Detection sensitivity: $r = \frac{TP}{FN+T}$.

Where TP is the number of MB found in the $\frac{\lambda}{4}$-neighborhood of a simulated MB. FP is the number of localized MB, but where no simulated MB can be found in a $\frac{\lambda}{4}$ -neighborhood. FN is the number of simulated MB where there is no localized MB to be paired within a $\frac{\lambda}{4}$-neighborhood.

### D. In Silico tracking metrics

To evaluate the performance of the tracking algorithm, we employed both local and global quality metrics. Local measurements involve assessing the accuracy of velocity estimates at specific points of interest, while global measurements involve evaluating the correlation coefficients calculated to quantify the relationship between the estimated velocities and the ground truth values.

$$\hat{C}(V_{GT}, V_{map}) = \frac{1}{N+1}\sum_{i=1}^{N}\left(\frac{V_{GT}^{i} - \mu_{GT}}{\sigma_{GT}}\right)\left(\frac{V_{map}^{i} - \mu_{map}}{\sigma_{map}}\right) \quad (5)$$

where $V_{GT}$ represents the velocity map obtained from the trajectories of the simulated positions and $V_{map}$ represents the velocity map obtained from trajectories found either by the LAT or TAL workflow. N is the number of pixels.

### E. In vivo metrics

Resolution was evaluated using the Fourier Ring Correlation (FRC) introduced for ULM in [56]. Briefly, an estimate of the image resolution is obtained by splitting the localization dataset into two sub-images and establishing a threshold for the consistency of the frequency information. To generate the two sub-images, all the trajectories obtained with the tracker were randomly separated into two datasets. The FRC was computed by using the spatial spectra $F_1$ and $F_2$ of each sub-image, as follows:

$$FRC(R) = \frac{\sum_{r\in R} F_1(r).F_2(r)^*}{\sqrt{\sum_{r\in R}|F_1(r)|^2.\sum_{r\in R}|F_2(r)|^2}} \quad (6)$$

The 1/2-bit threshold was used to assert image resolution, which corresponds to the highest spatial frequency containing at least 1/2 bit of information [57].

### F. In Silico validation dataset

The *in silico* PALA dataset is composed of 20 buffers with 1000 simulated ultrasound frames, each frame containing an average of 24 moving MB at various velocities in crossing vessels of varying sizes and shapes [47]. To achieve artificially higher MB concentration, we implemented a technique in which the complex IQ images of consecutive buffers were summed frame by frame. We consider the concentration was constant between the buffers. This process resulted in the creation of buffers with an identical frame rate, yet with an increased number of MB per frame. This approach is based on a simple argument of linearity. Since the simulations that were used are linear, it is exact to compound buffers to generate additional MB. By doing so, the number of buffers was reduced while preserving the same total number of simulated MB. The concentration (C) is presented in arbitrary units based on the number of buffers summed. C=5 a.u. means that 5 consecutives buffers have been summed to form a buffer five times more concentrated than the original ones. Note that since the total number of MB is maintained, higher concentrations are associated with shorter acquisition times. For the different concentrations, the same parameters as those shown in Table I are used.

### G. In vivo validation dataset

The *in vivo* PALA dataset is composed of 240 buffers, containing 800 frames at 1000 frames per second, after compounding of rat brain measurements each. A craniotomy was performed and MB were injected at 80 $\mu L.min^{-1}$ during 4 min [47]. As in the *in silico* study, the concentration was artificially increased while maintaining a constant total number of MB, before the application of the SVD. Higher concentrations are thus associated with proportionally shorter acquisition times. For the different concentrations, the same parameters as those shown in Table I are used.

*Table I* TAL and LAT/LATs parameters for the different datasets.

| Dataset | LAT & LATs | TAL |
|---|---|---|
| *In Silico (beamforming grid: λ)* | Nb = 90<br>MLD = 2pixels<br>MFG = 0 frame | τ = 0.5<br>$\epsilon = 1.4$ |
| | $N_f = 15$ | |
| | | $f_c = 45Hz$ |
| Rat ULM *(beamforming grid: λ)* | Nb = 300<br>MLD = 2 pixels<br>MFG = 0 frame | τ = 0.4<br>$\epsilon = 1.4$ |
| | $N_f = 15$ | |
| | | $f_c = 45Hz$ |
| Rat and Mouse DULM *(beamforming grid: λ/4)* | Nb = 500<br>MLD = 4 pixels<br>MFG = 4 frames | τ = 0.4<br>$\epsilon = 1.4$ |
| | $N_f = 20$<br>$f_c = 45Hz$ | |

### H. In vivo Dynamic ULM rat and mouse datasets

The acquisitions performed for the DULM study adhered to the guidelines outlined in the "Guide for the Care and Use of Laboratory Animals" set forth by the Canadian Council for Animal Care. The rat study procedures were granted ethical approval by the Animal Care Ethics Committee of the Montreal Heart Institute (Permit Number: 2019-2464, 2018-32-03). Similarly, the mouse study procedures were approved by the McGill University Animal Care Committee under the regulations of Animal Use Protocol AUP-4532.

#### 1) Rat model

The surgical procedure and ultrasound acquisitions were performed on a 500 g female adult rat under 2% isoflurane anesthesia. Throughout the procedure, the body temperature was maintained at 35°C using a small animal monitoring platform (Labeo Technologies Inc., QC, Canada). To minimize movement, the rat's head was secured in a stereotaxic frame, and a surgical micro drill was used to remove the skull, creating a window over the brain measuring approximately 15 mm × 10 mm.

Three datasets were obtained for the same two-dimensional plane using different concentrations. Prior to ultrasound acquisitions, an intravenous bolus injection of a MB solution ($1.2 \times 10^{10}$ MB per milliliter, Definity, Lantheus Medical Imaging, MA, USA) was administered through the jugular vein. The injection lasted 30-60 seconds and consisted of 50, 25 and 12.5-μL MB solutions diluted in saline, followed by a saline flush using a 27G needle.

A Vantage system (Verasonics Inc., Redmond, WA) emitting 3-cycle pulses centered at 15.625 MHz using 3 tilted plane waves (-1°/0°/1°) was used to sample 384 radiofrequency (RF) buffers with a 100% bandwidth sampling scheme, each lasting 400 ms at 1000 frames per second, after compounding (153 600 frames in total for each acquisition). The Mechanical Index (MI) of the sequence was calculated in a water tank at the focal of the probe and it's derated value is equal to 0.03. Each buffer was gated on the R-wave of the ECG recorded using the monitoring platform.

*2) Mouse model*

The ultrasound acquisitions were conducted on a 7-week-old male mouse under 2% isoflurane anesthesia. The mouse's body temperature was maintained at 35°C using a feedback heated blanket. To minimize movement, the head of the mouse was secured in a stereotaxic frame. The ultrasound acquisitions started following a tail vein bolus injection of 80uL MB solution. (Freshly activated Definity MB were diluted in sterile PBS at 7.4pH in a 1:10 ratio immediately before injecting). Each buffer was gated on the R-wave of the ECG recorded using the monitoring platform.

3-cycle pulses centered on 15.625 MHz using 11 tilted plane waves (-5° to 5° with 1° step) were used. We acquired 400 RF buffers, each lasting 600 ms, at a rate of 1000 frames per second after compounding, using a 100%-bandwidth sampling scheme (240 000 frames in total). The MI of the sequence was calculated in a water tank at the focal of the probe and is equal to 0.1. Note that these acquisitions were transcranial, so the effective MI is likely to be smaller.

## III. Results

### A. In Silico localization performance of TAL

*1) Performance at different concentrations*

Figure 2 presents both qualitative (Fig. 2b) and quantitative (Fig. 2c) comparisons between the standard LAT, LATs workflows and the proposed TAL workflow for reconstructing anatomic vessels *in silico* at various MB concentrations. The ground truth density map (Fig. 2a) is formed by accumulating all simulated MB positions. The qualitative performances for different concentrations (Fig. 2b) are presented within a selected region of interest corresponding to a challenging horseshoe pattern. The quantitative performances (Fig. 2c) are assessed using four different metrics: localization error, detection rate, precision, and sensitivity.

Across all concentrations, TAL consistently achieved a smaller localization error (maximum of 0.18 λ) compared to LAT's smallest localization error, which was 0.23 λ. For the standard concentration (C=1 a.u.), the localization error of TAL was measured at 0.14 λ, whereas the LAT pipeline exhibited a larger localization error of 0.23 λ. Adding the cubic smoothing spline to the LAT pipeline (i.e., LATs) drastically improved performance. At the standard concentration (C=1 a.u.) the localization error of LATs is 0.15 $\lambda$ which almost matches TAL's error of 0.14 $\lambda$.

For the same concentration (C=1 a.u.), TAL achieved a detection rate of 60%, while LAT achieved a detection rate of 27% and LATs a detection rate of 54%. At the extreme concentration (C=10 a.u.), all methods exhibited poor detection rates, with TAL at 14%, LAT at 2% and LATs at 3%.

The detection precision represents the ratio of true positions found to all detections made. For the standard concentration (C=1 a.u.), TAL and LATs achieved a precision of 85% and 82% indicating approximately seven times more true detections than false ones. In contrast, LAT achieved a precision of 50%, indicating an equal number of true and false detections. As the concentration increased, only TAL and LATs consistently maintained a precision rate above 50%.

The detection sensitivity reflects the proportion of MB found in relation to the total number of simulated MB that can be found. TAL demonstrates a sensitivity of 67% at the standard concentration (C=1 a.u.), LATs a sensitivity of 62% while LAT achieved a sensitivity of 37%.

We also note that as the concentration increases, the disparity between LATs and TAL widens in favor of TAL across all metrics. In addition, all TAL metrics at a doubled concentration (C=2 a.u.) were superior to LAT metrics at the standard concentration.

*2) Performance at different SNR*

Figure 3 quantitively illustrates the robustness of TAL, when exposed to an increasing noise in the IQ data at the standard concentration (C=1 a.u.). At high SNR, both LAT and TAL exhibited comparable performances, although TAL appeared to yield incrementally better results. Notably, some TAL metrics degraded starting at approximately 25 dB, whereas all four LAT metrics degraded starting at approximately 40 dB. LATs exhibited a similar trend to TAL as the concentration increases, although TAL performs better at low SNR.

Specifically, for TAL, the localization error remained stable between 50 and 20 dB of SNR, while the localization error was increasing as the SNR decreases for LAT. At high SNR, both approaches demonstrated a comparable RMSE.

The detection rate indicates that the three methods performed similarly well at high SNR (50-40 dB) and low SNR (15 - 10 dB). However, in the SNR range between 40 dB and 15 dB, TAL and LATs achieved better performances although TAL's performance drops off later (20 dB) compared to LATs

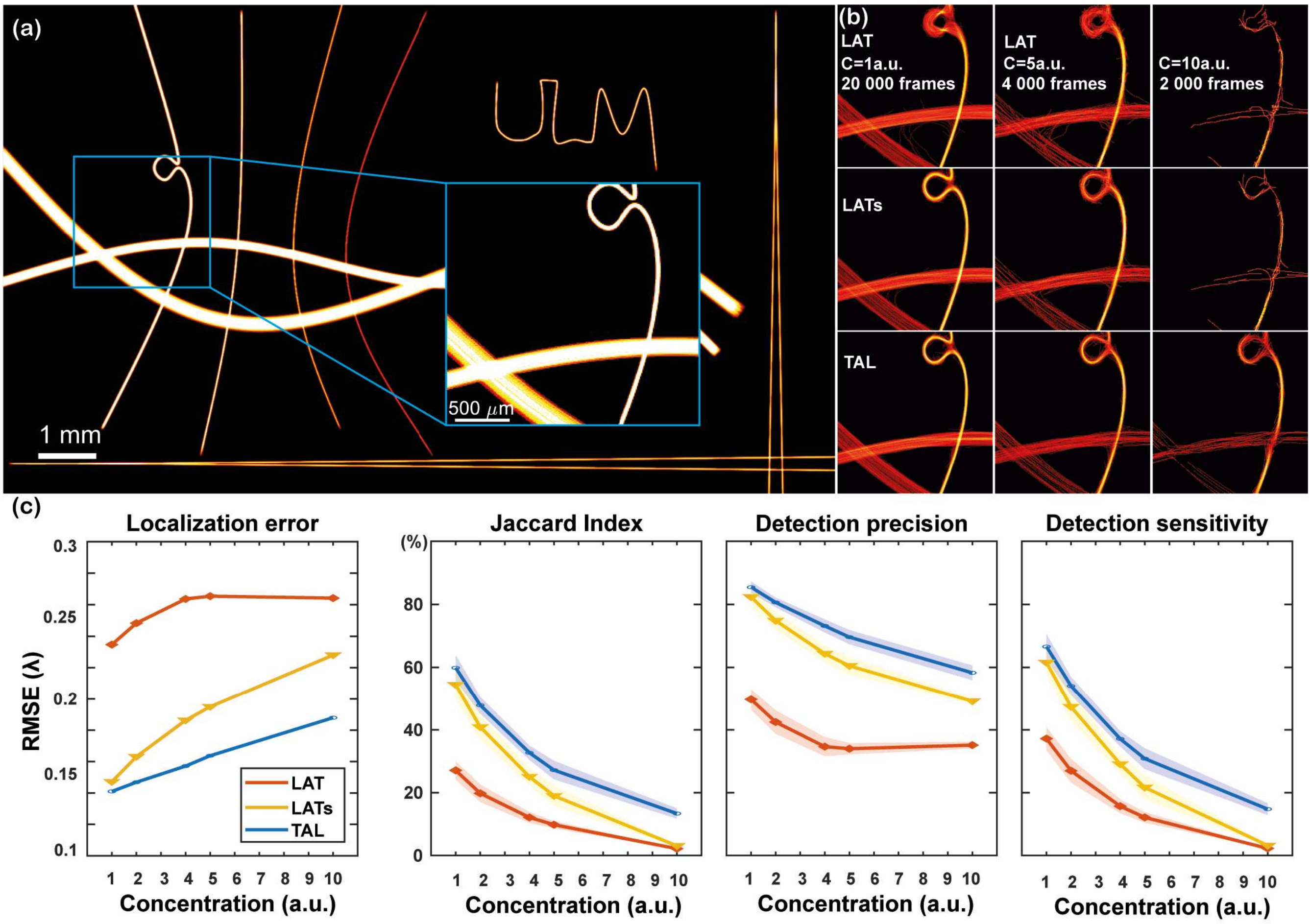

*Figure 2* **In silico localization and detection performances at different concentrations**. (a) Ground truth density map resulting of the accumulation of positions in 20,000 frames divided into 20 buffers. (b) Zoomed-in renderings for LAT, LATs and TAL pipelines at different MB concentrations at a constant Signal-to-Noise Ratio (SNR) of 20 dB. (c) Performance metrics: localization error and statistical measurements, including the detection rate (Jaccard index), precision, and sensitivity, for LAT, LATs and TAL pipelines, at a fixed Signal-to-Noise Ratio (SNR) of 20dB. The statistical measurements were computed for each buffer and shown as mean values with their respective standard deviations.

(25 dB). At 30 dB, 25 dB, and 20 dB, the differences in detection rates between LAT and TAL were 7%, 17%, and 33%, respectively.

The sensitivity for the three methods is following the same trend as the detection rate.

The precision analysis revealed that from an SNR of 60 dB to 15 dB, TAL maintained similar rates, indicating that the increased noise did not lead to a higher proportion of false detections relative to true detections. The precision rates for TAL were 86% at 50 dB and 84% at 15 dB. In contrast, the LAT algorithm exhibited a rapid decline in precision as SNR decreased, with rates of 83% at 50 dB and 28% at 15 dB. Adding the spline in LATs reduced the proportion of false positives, resulting in performance comparable to TAL, although TAL's performance remained more consistent.

*3) Performance for different parameters: $\tau$ and $\epsilon$*

Figure 4 provides a quantitative assessment of TAL's robustness to parameter variations across different noise levels in the IQ data, focusing on the standard concentration (C=1 a.u.). Notably, changes in the $\tau$ parameter within the range of 0 to 1 do not notably impact the method's performance. Conversely, the $\epsilon$ parameter, with values exceeding 1, exhibits more sensitivity to parameter adjustments, particularly for extreme values. Interestingly, the results stabilize when $\epsilon$ slightly exceeds 1, but deteriorate with extreme $\epsilon$ values.

### *B. In Silico tracking performance of TAL*

Figure 5 presents the tracking performance globally and locally for the different methods via the velocity estimation for different concentrations. All the simulated MB trajectories were derived to obtain MB velocities. Those velocities were averaged at the positions of the associated MB to form the ground truth velocity map (Fig. 5a).

The global performance of the tracking is reflected by the correlation coefficient between the ground truth velocity map and the velocity maps obtained from the tracking of each method (Fig. 5b). At the standard concentration (C=1 a.u.), TAL achieved a correlation rate of 90%, whereas LAT and LATs achieved 78% and 85%. Similarly, at 5 times the standard concentration (C=5 a.u.), TAL achieved a correlation rate of 75%, which is comparable to LAT at the standard concentration (C=1 a.u.). Adding the spline in LATs increased the correlation compared to LAT at low concentration. Nevertheless, beyond five times the standard concentration, LAT and LATs performed similarly. All three methods exhibited a degradation in the velocity estimators as the concentration increased.

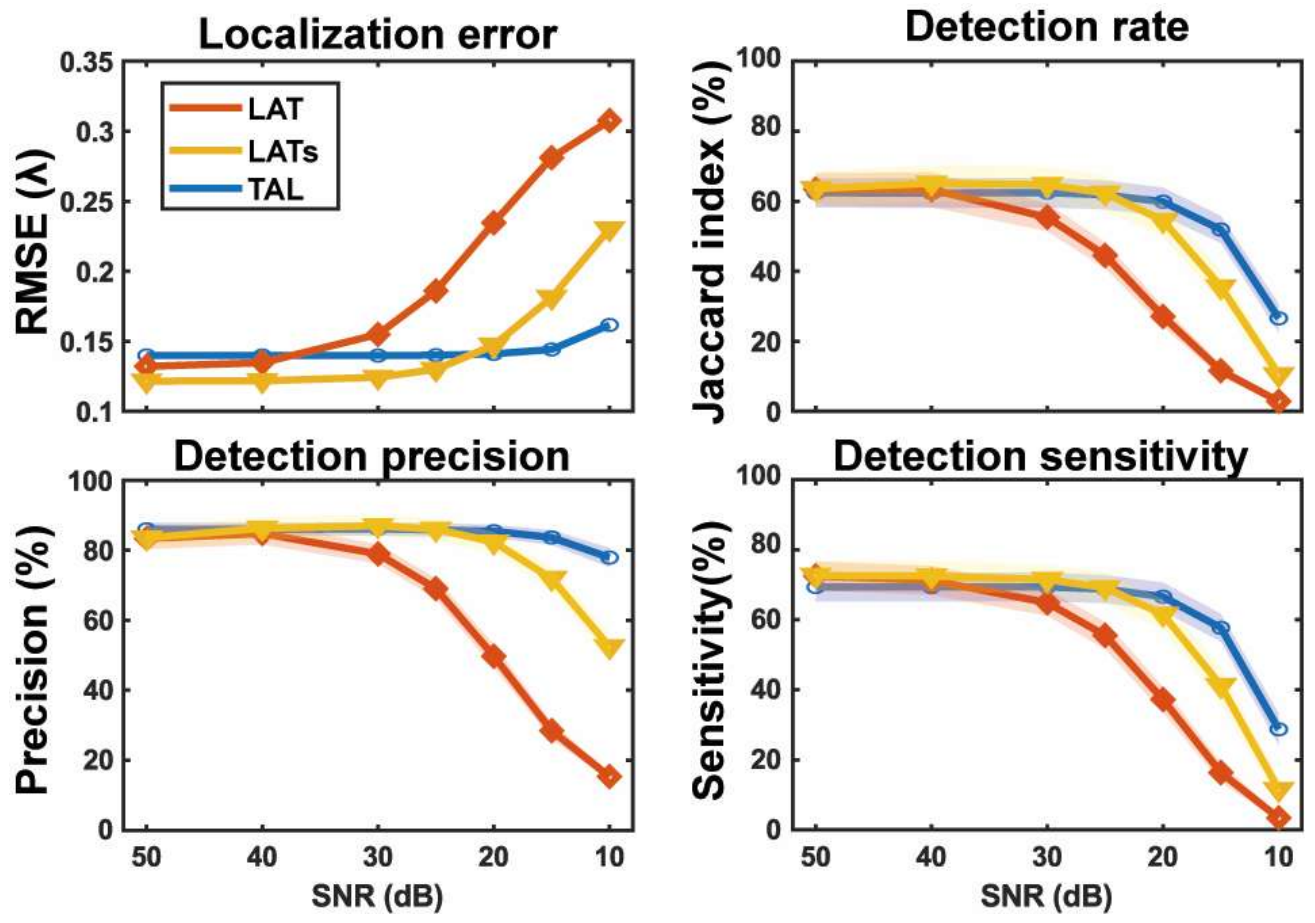

*Figure 3* **Noise localization and detection robustness.** The four different metrics are presented for the standard concentration (C=1 a.u.). The statistical measures are computed across different numbers of buffers and shown as mean values with their respective standard deviations.

Locally, velocity profiles orthogonal to the flux in two distinct areas of the image were extracted (Fig 5c). Those two profiles are indicated by the pink and green lines in Fig. 5a. In the pink profiles, where both methods must estimate high velocities in large vessels, TAL demonstrated a closer estimation of the velocity than LAT & LATs as well as a higher robustness in the estimation for the different concentrations, although the estimator was degraded as the concentration increased. We can notice a bias for the velocity positions of the vessels obtained with LAT, which is corrected by adding the cubic smoothing splines in LATs. In the green profiles, where low velocities in a small vessel are estimated, TAL provided a better estimation of velocity at the standard concentration (C=1 a.u.). As concentration increased, all methods exhibited less accuracy, but TAL appeared to be more consistent than LAT & LATs.

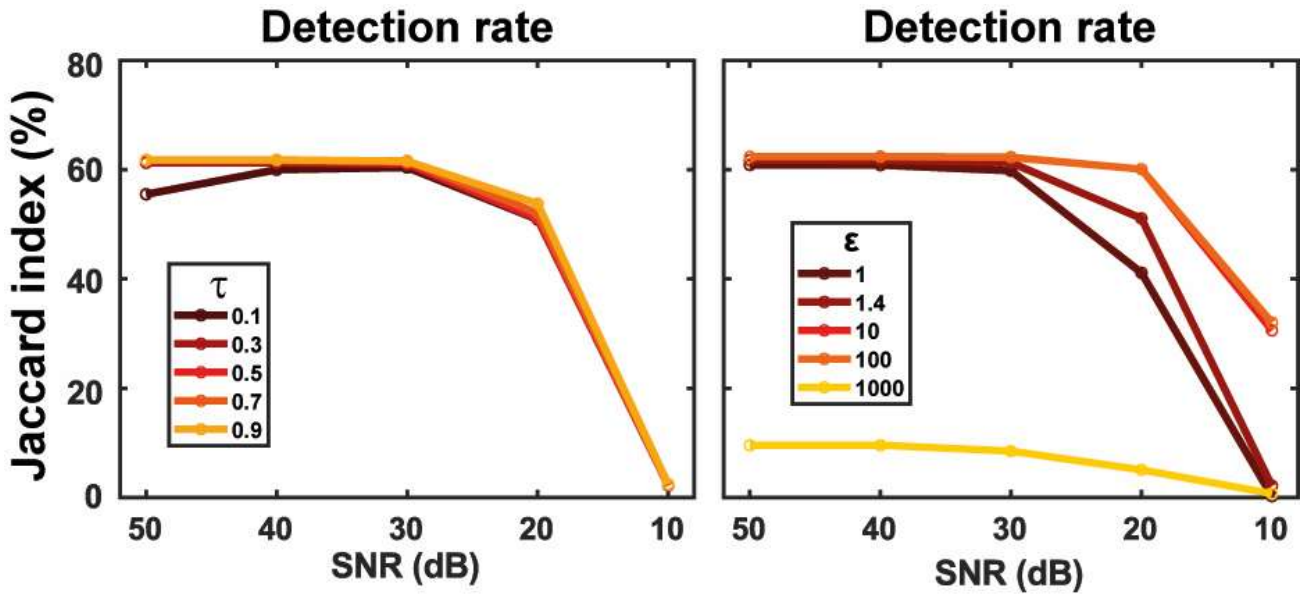

*Figure 4* **TAL robustness across varied τ and ε value.** The Jaccard index is presented for the standard concentration (C=1 a.u.) and for different SNR.

## C. *In vivo performance of TAL for ULM*

Figure 6 displays the different in vivo density maps obtained for various concentrations using the different methods ($C \in [1,5,10]\ a.u.$, see Fig. 6a) and closeup patches as well as normalized intensity profiles (Fig. 6b). The first two rows illustrate the density maps obtained with the LAT and LATs workflows, while the third row presents the density maps obtained with the TAL workflow.

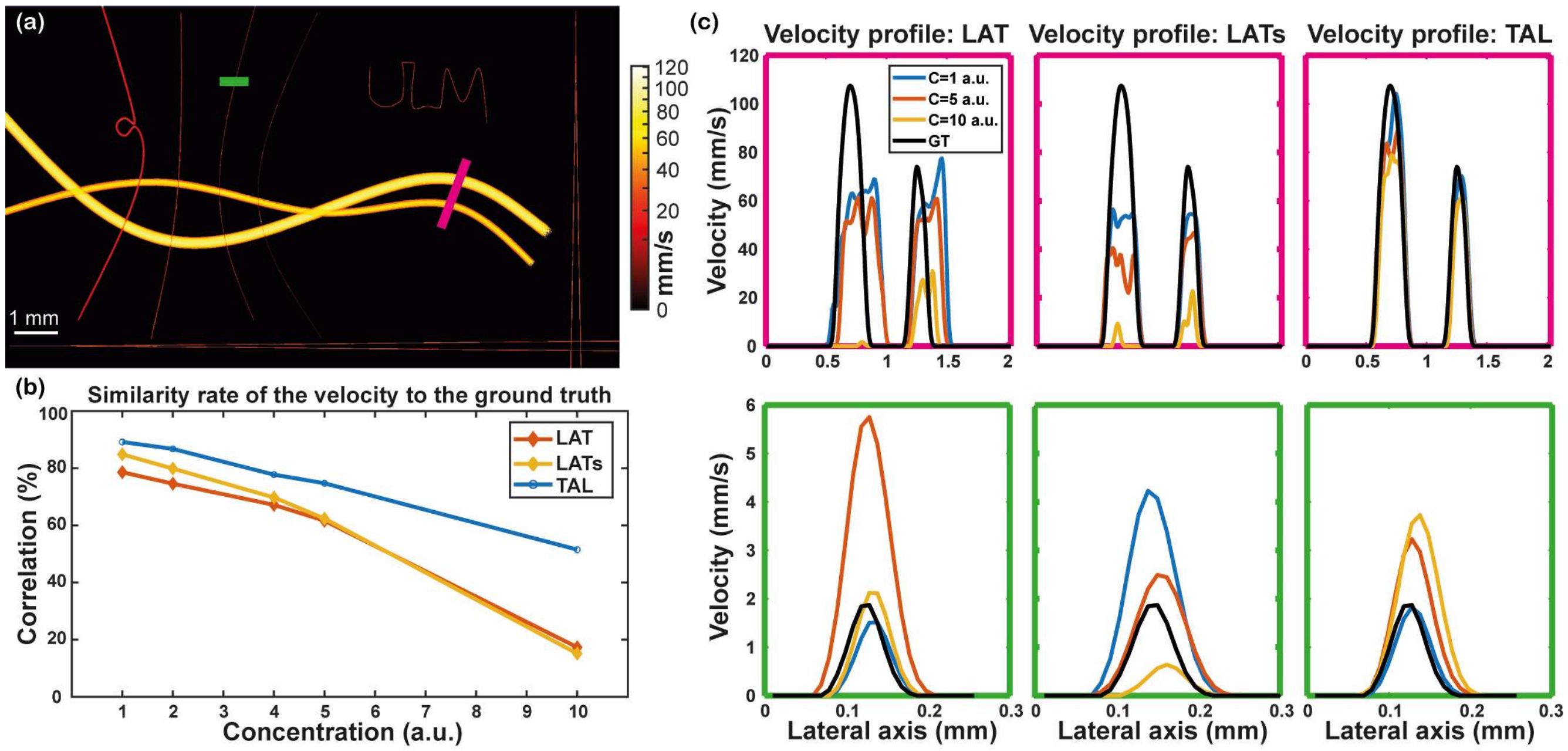

*Figure 5* **In Silico performances of the velocity estimator at different concentrations.** (a) Ground truth velocity map resulting of the accumulation of the velocity at the positions of the MBs in 20,000 frames divided into 20 buffers initially. (b) Correlation coefficient between the ground truth velocity map and the velocity maps obtained from each approach. (c) Velocity profiles along two different lines for LAT, LATs and TAL workflow, each line taken orthogonal to the vessels to visualize the Poiseuille velocity profiles modeled in this simulation. (d) Superposed velocity profiles of the ground truth and both methods for different concentrations.

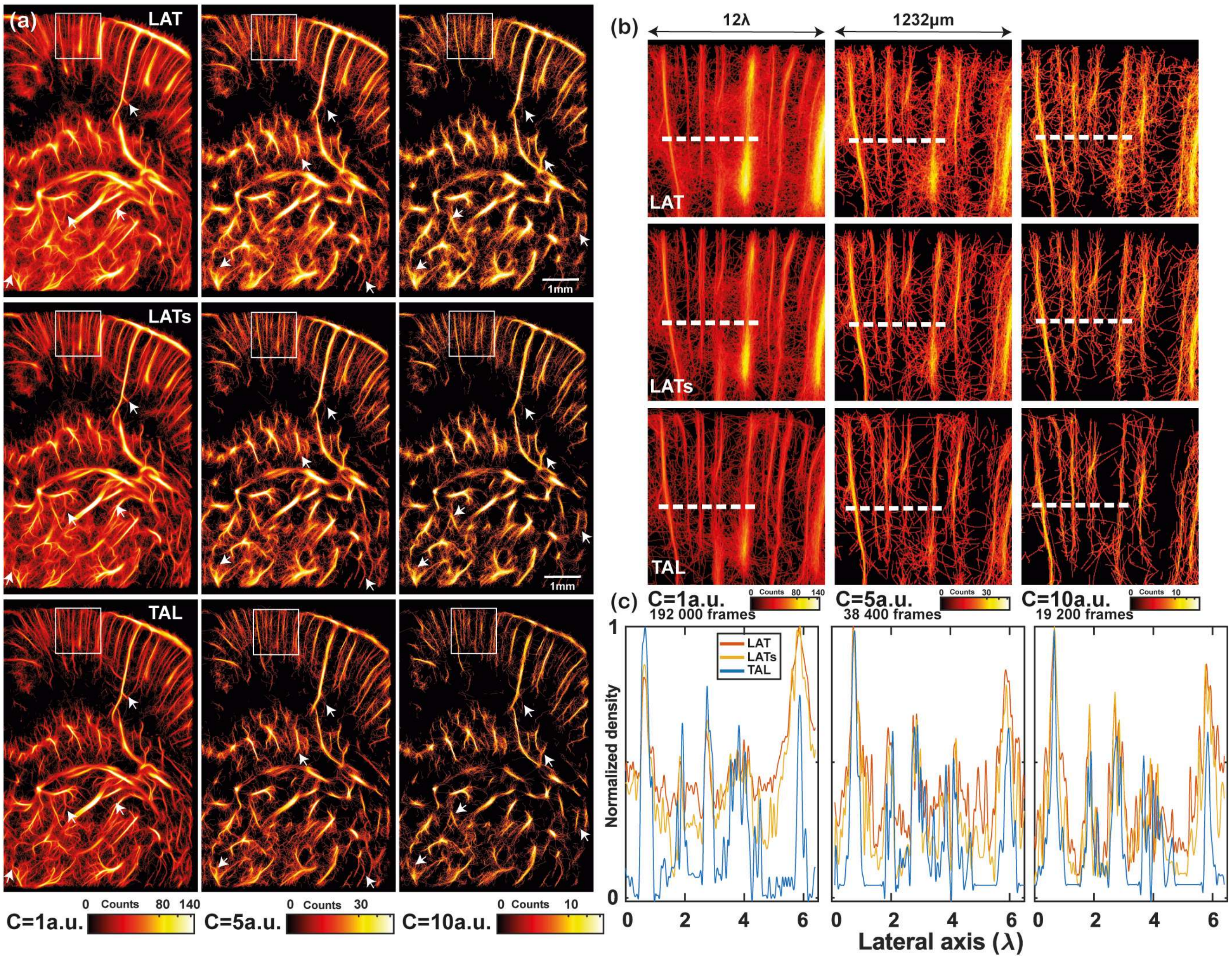

*Figure 6* **Impact of concentration on the vessel rendering in vivo.** (a) First row presents the different density maps obtained with the standard approach (LAT pipeline); the second row present the density maps obtained with the LATs approach while the third presents the density maps obtained with the proposed method (TAL). Each density map is compressed with a cubic rate and saturated with a factor 0.5 relative to the maximum intensity. (b) Zoom-in of ULM angiograms on the cortex region (white square in Fig. 5a) at different concentrations for the different approaches. (c) Normalized intensity profiles along a given horizontal line overlaid by the white dotted line.

Qualitatively, there was a similarity between the two methods for the standard concentration, even if we notice some regions (pointed with the white arrows) where TAL recovered more details. However, as the concentration increased, while both methods detected fewer vessels, TAL detected more than LAT, as illustrated by the white arrows. Moreover, as the concentration increased, both the number and length of detected vessels decreased. Adding the splines in LATs enhanced the differentiation of the vessels compared to LAT.

The associated FRC for each approach, at different concentrations, was also calculated (see Table II). At the standard concentration, TAL achieved a resolution of 21.3-µm, while LAT and LATs achieved resolutions of 24.9-µm and 22.9-µm, respectively. As the concentration increased, the resolution decreased for all approaches. TAL maintained an incrementally better robustness to resolution - e.g., at a five-fold increase in concentration, TAL, LAT and LATs achieved a 27.7-µm, 39.9-µm and 31.2- µm resolutions, respectively.

*Table II* In vivo resolution performance (FRC) of ULM angiogram reconstruction at different concentrations for LAT, LATs and TAL workflows

| Method | C=1 a.u. | C=2 a.u. | C=5 a.u. | C=10 a.u. |
|---|---|---|---|---|
| LAT | 24.9 $\mu m$ | 28.1 $\mu m$ | 32.9 $\mu m$ | 38.9 $\mu m$ |
| LATs | 22.9 $\mu m$ | 26.7 $\mu m$ | 31.2 $\mu m$ | 36.9 $\mu m$ |
| TAL | **21.3 $\mu m$** | **22.9 $\mu m$** | **27.7 $\mu m$** | **32.3 $\mu m$** |

From the normalized density profiles extracted along the white doted lines in the zoom-in regions in the cortex (Fig. 6b), we can see that for the two lowest concentrations, the differences in intensity between the vessels and the background are larger for TAL.

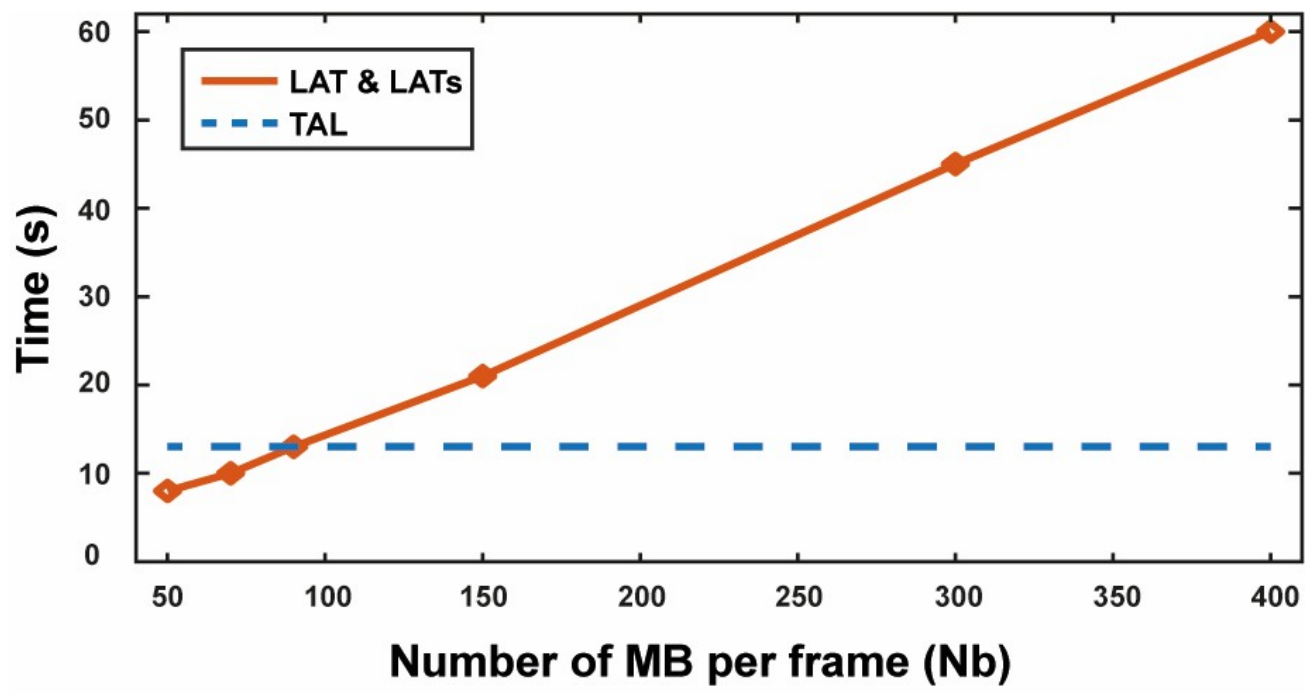

*Figure 7* **In vivo processing time** *of* 1 buffer (78×118×800) for LAT & LATs and TAL workflows with varying input parameter Nb (number of MB per frame).

Both methods showed robustness to MB concentration in this region, as the number of vessels detected is similar.

Figure 7 provides an overview of the computational cost associated with both methods. Notably, the computational cost of the TAL approach is solely contingent upon the size of the volume data, whereas that of the LAT/LATs approaches hinge on the number of MB considered. Upon examination of the *in vivo* datasets, it becomes apparent that the TAL approach outpaces the LAT/LATs approaches in terms of speed if $N_b >$ 90.

### D. *In vivo performance of TAL for DULM*

#### 1) *Performance at different concentrations*

Figure 8 presents qualitative differences in the left hemisphere signed density maps of a rat brain for concentrations of 12.5 µL and 50 µL obtained using LATs and TAL (Fig. 8a). Quantitative differences are depicted when velocity extraction is performed over time for representative vessels using both approaches at concentrations of 12.5 µL, 25 µL and 50 µL (Fig. 8b). Qualitatively, some vessels showed differences in radius between concentrations, as exemplified by the vessel pointed out by the red arrow. At depth, both small and big vessels as well as connections between vessels were lost at high concentration. The arrows point to vessels where the velocity profiles along two cardiac cycles have been extracted in the Fig. 8b. A difference in velocity amplitude can be noticed between both methods; for instance, TAL can detect a pulsatile flow, with two clear local maxima waveform patterns, where they are expected, in all three vessels while it is less clear in the case of LATs.

#### 2) *Intra-animal comparison*

Figure 9 shows a signed ULM reconstruction in a mouse brain using LATs and the proposed methods: TAL, with vessels highlighted by yellow, blue, red, and green arrows (Fig. 9a). Qualitatively, differences can be observed between the two approaches: overall, TAL retrieved more vessels, particularly in the cortex region close to the superior sagittal sinus, where LATs could not retrieve the vessels identified by TAL. The velocities of the highlighted vessels over 5 complete cardiac cycles were extracted as shown in Fig. 9b for both methods. For analysis, those vessels correspond to pairs of symmetrical vessels between the hemispheres. It is observed that the mean velocity in both hemispheres is similar (solid line for the left hemisphere and dotted line for the right one) for both methods, but TAL exhibits more consistent and synchronized fluctuations during diastole and systole events (Fig. 9b).

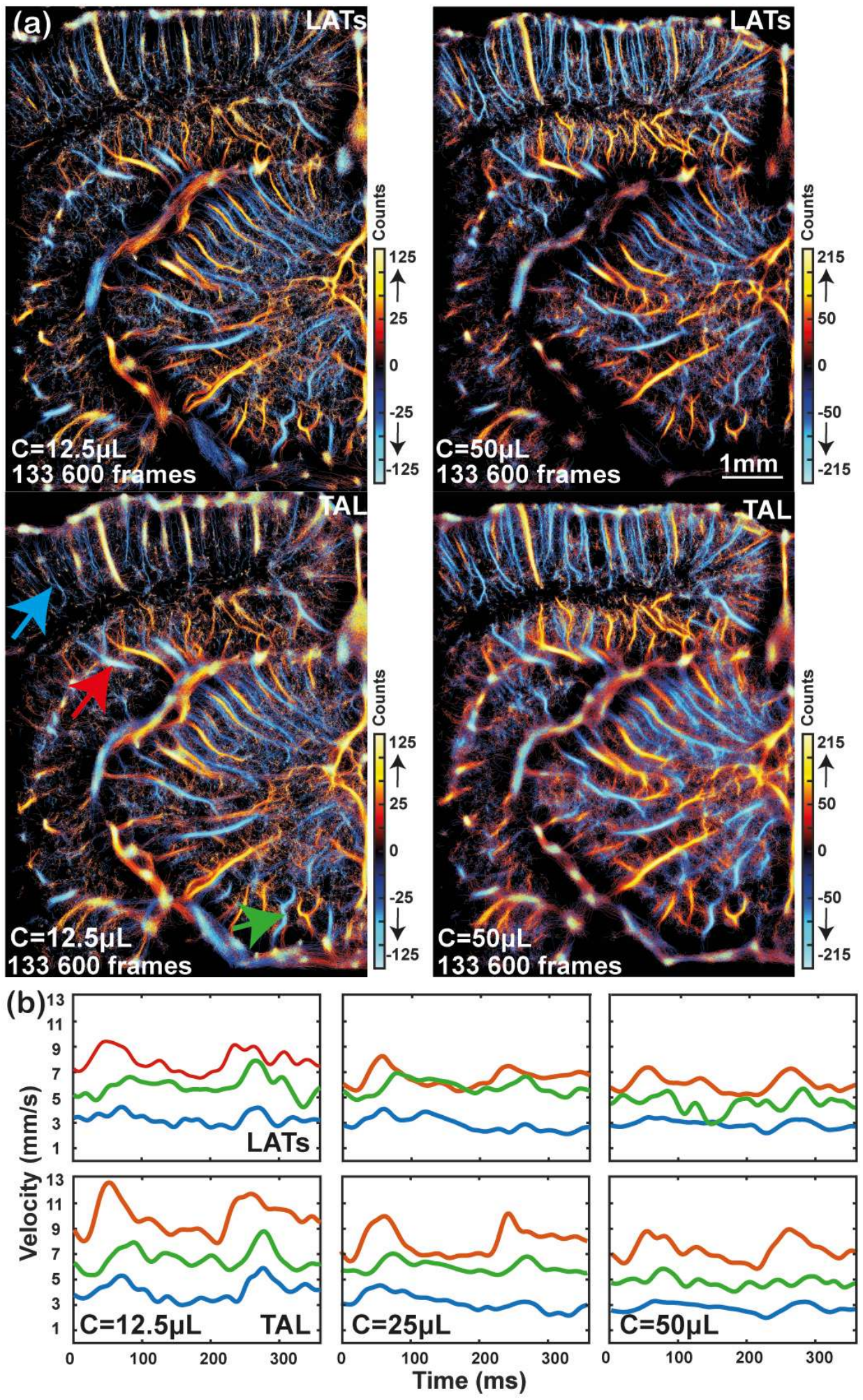

*Figure 8* **Velocity estimator comparison for DULM application in vivo at different concentrations in a rat brain.** (a) Signed ULM maps in a rat brain for two different concentrations of microbubble (C=12.5µL and C=50µL) obtained for the first row with the LATs pipeline and for the second row with the TAL pipeline. (b) Velocity profiles for different vessels overlaid by different colored arrows (blue, red, and green) from the lowest concentration to the highest one (left to right) (12.5µL-25µL-50µL) obtained with the LATs and TAL methods.

## IV. Discussion

In this study, we introduced a processing approach that tracks prior to localizing the MB in ULM and DULM. The idea is related to previous studies implicitly modelling the spatio-temporal context with deep learning to better disambiguate MB trajectories at high concentration in 2D [27], [40], [58].

Specifically, we aimed to address the challenge of finding and tracking a larger number of MB in high concentration environments. To do so, we sought to replace the tracking part with a physically motivated pairing approach that considers global information from all frames in a buffer, rather than relying on frame-to-frame pairing based on minimal distance between localizations in adjacent frames. The choice of a tracking approach is particularly important, as recent publications in the field emphasize studying *in vivo* velocity to extract biomarkers [11], [12], [16], [28], [59]. However, velocity estimation relies on the derivative of the trajectories, making it essential to have an accurate tracker in addition to an efficient space detection and localization algorithm to accurately estimate local blood flow. Nevertheless, the method relying on minimum distance, even in scenarios with high frame rates, remains highly susceptible to generating false trajectories due to the possibility of missing detections in certain frames and the failure to account for the continuity of movement, especially in high concentration environments. To demonstrate these objectives, we benchmarked the TAL method against the state-of-the-art LAT pipeline, both *in silico* and *in vivo*.

In the *in silico* experiments, we demonstrated that TAL yields superior reconstruction maps using four metrics at any concentration multiplier ranging from 1 to 10 (see Fig. 2c). Specifically, we showed that at twice the concentration, TAL outperformed the LAT workflow at standard concentration. These results align with our first objective, as TAL allows for a twofold increase in concentration, potentially reducing acquisition time while outperforming the standard method. The RMSE for the TAL approach increases linearly with concentrations while the LAT RMSE does not. This phenomenon could be attributed to the rejection of detections in the LAT approach when there are more than two MB in the vicinity around the detection. Consequently, only isolated MB are localized in LAT. In contrast, TAL applies the localization algorithm on each detection individually, leading to an expected increase in error as the likelihood of superposed MB signals rises with increasing concentrations. Moreover, the cubic smoothing spline improves the RMSE by adding temporal regularization from the tracking. The steeper slope of LATs compared to TAL may indicate less effective tracking by LAT as the concentration increases. Additionally, we observed that TAL exhibits robustness to increased noise levels, as evidenced by its performance across varying SNR levels *in silico* (see Fig. 3). The enhanced robustness observed in comparison to the LAT method may stem from differences in MB detection mechanisms. In LAT, MB are identified based on the highest intensity peaks, which can lead to issues in noisy acquisitions where noise might be mistaken for MB, or actual MB could be obscured by noise. Conversely, in the TAL approach, emphasis is placed on the geometry in space-time rather than intensity. As a result, the method is potentially less likely to detect incoherent noise that occurs sporadically in time, thereby preserving the integrity of MB trajectories. This robustness appears to contribute to the reproducibility of results between experiments, where SNR can significantly vary due to experimental factors. By studying velocity *in silico*, we revealed the biases that arise in velocity estimators based on the chosen method and varying concentrations (see Fig. 5). TAL showed better efficiency in tracking MB and extracting accurate blood flow velocity. Specifically, we demonstrated that for low velocities (highlighted in the green box in Fig. 5), the velocity estimated by the TAL approach exhibits a notably better standard deviation. Another point of interest is the shift of the velocity peak that can be attributed to the linear interpolation utilized in the LAT approach. Indeed, by adding splines in LATs, the shift is all but eliminated.

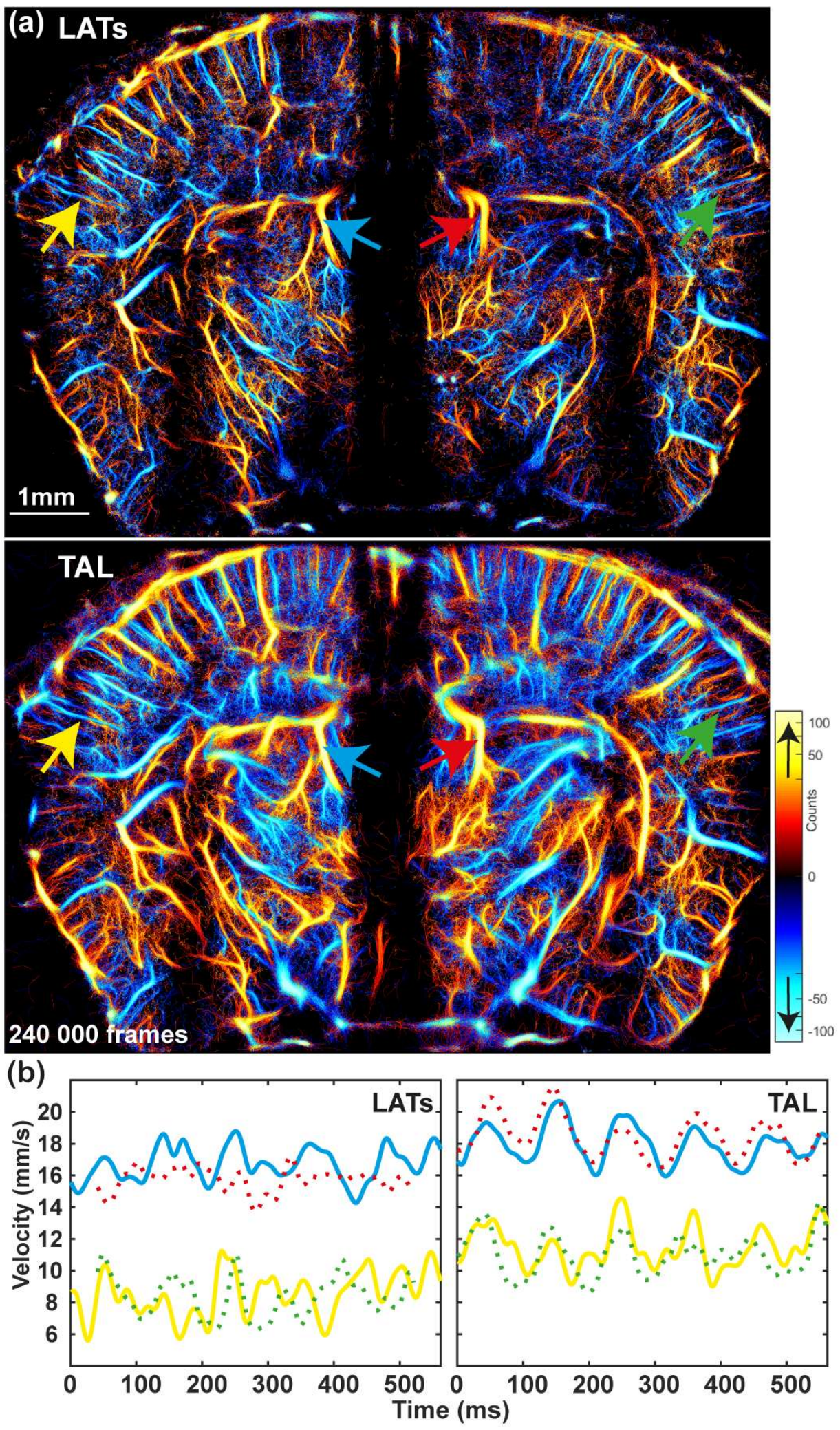

*Figure 9* **Hemodynamics coherence between left and right hemisphere in a mouse brain.** (a) Signed Ultrasound Localization Microscopy map of a mouse brain obtained for the first row with the LATs pipeline and for the second row with the TAL pipeline. (b) Comparison between left and right hemisphere (solid lines: left hemisphere, dotted lines: right hemisphere) of velocity profiles for different sizes of vessels obtained with the LATs & TAL methods.

We chose to enhance the state-of-the-art LAT by adding a cubic smoothing spline, to match the type of smoothing used in TAL, resulting in a new approach called LATs. Incorporating the spline improved all in silico performance metrics of LAT and corrected the shift of the velocity peak observed in Figure 5. However, TAL still outperformed LATs, indicating that the cubic smoothing spline alone does not account for TAL's superior performance. Consequently, we decided to conduct the DULM processing with LATs instead of LAT.

To further validate our findings, we conducted *in vivo* validation using an open-source rat dataset and demonstrated that TAL is more robust to increasing concentration with associated reduction of acquisition times (see Fig. 6a) and maintains better resolution in the FRC study (see Table II). Density profiles also confirmed better resolved vessels in the cortex (see Fig. 6b). Indeed, at lower concentrations, the background intensity between vessels is lower with TAL, reflecting its ability to highlight more likely MB trajectories. In fact, TAL only enhances positions that are susceptible to belong to a trajectory, while LAT enhances all the positions that are likely to be a MB in space by identifying the highest intensity peaks in the B-mode.

Another point of interest in the cortical zoom is the absence of numerous small vessels between the vertical vessels with the TAL approach compared with the LAT approach. These small vessels are used for many ULM biomarkers [16] but are difficult to characterize with certainty as vessels. It is important to emphasize that our proposed method remains highly sensitive. In other regions of the brain, we observe a greater detection of vessels compared to the LAT approach. Additionally, through simulation studies, we have demonstrated that the TAL approach is less prone to generating false detections. We also successfully applied TAL to different animal models, including rats and mice.

We applied LATs and TAL to DULM and observed consistent velocity estimations across different acquisitions and concentrations within the same animal for both methods (see Fig. 8 and 9). In DULM, the TAL approach seems to extract a better pulsatile blood flow during the cardiac cycle. Between the three rat datasets, the blood flow differences observed could be attributed to the increasing concentration biasing the tracking process as illustrated *in silico*. It could also be attributed to physiological changes in heart beating caused by prolonged anesthesia, as the datasets were acquired successively. Additionally, we demonstrated the coherence of both methods in estimating mean velocities by comparing symmetrical vessels in both hemispheres of a mouse (see Fig. 9). TAL results seem to indicate more accurate velocity variations caused by the heartbeat of the animal. The cubic smoothing spline ensures the continuity on velocity and acceleration compared to the linear interpolation. In addition, the choice of a cut-off frequency leads to a better control over the smoothing.

It is however important to acknowledge certain limitations. In the *in silico* experiment, while the TAL approach yields generally superior results across various metrics, it is noteworthy that at a concentration of C=5 a.u., the bottom vessels may appear less bright compared to LAT. This local disparity could potentially be minimized given the very high concentration involved (120 MB on average per frame). In the *in vivo* experiments, the ground truth for microvessels structure and blood flow velocity remains unknown. To validate our angiography and velocity estimation, it would be necessary to compare it with another imaging technique that can obtain similar spatial and temporal resolution, such as optical coherence tomography at superficial depth (1 or 2 mm). While we highlighted issues with velocity estimations caused by the tracking algorithm *in silico*, it's important to note that the simulations were limited in representing the complexity of brain vasculature. Therefore, the conclusions drawn from the *in silico* experiments may not necessarily generalize to *in vivo* scenarios. Realistic simulations of brain vasculature could point out other and more specific issues [25]. In our effort to minimize parameter modifications across different concentrations, it is conceivable that the parameters employed in the standard LAT or TAL methodologies may not be optimal for each concentration under investigation. While there might be specific regions where LATs could perform better than TAL, overall, TAL consistently achieved better results. The use of the Jerman function could produce artificial trajectories in principle, but because we can prevent those detections by enhancing only structures with the diameter of the PSF and structures that are coherent in time during a fixed number of frames, we think it is less likely to occur according to the detection precision metrics in Fig. 2. The thinning algorithm applied on the binarized Hessian volume benefits from a grid thinner than $\lambda$. Otherwise, if the width of the tubular structure (i.e., the spatiotemporal PSF MB) is reduced to 1 pixel, the thinning algorithm will not be effective, as the structures are already too thin. This could potentially result in numerous detection failures. Finally, the method is based on the continuity of the MB signals between frames, so it is particularly well suited for high frame rate imaging. The MB with the lowest velocities are the easiest to track if they have not been removed by the SVD, and if the velocity of MB are too important compared to the frame rate the trajectories would look like a pearl necklace in the spatiotemporal domain and not a tubular structure anymore and may not be enhanced by the method.

From a processing perspective, the proposed method offers several advantages. First, it requires only three parameters to set, $\tau$, $\epsilon$ and the length of the tracks $N_f$ (see Table I), $f_c$ being an interpolation parameter. In particular, the Fig. 4 shows that in addition to few parameters, the influence of those parameters on the results are minimal. *In vivo*, we employed a $\epsilon$ value of 1.4 across all datasets, acknowledging that it may not be the optimal parameter choice. However, this highlights that TAL does not require meticulous adjustment of parameters to achieve the reported results in the article. Additionally, the calculation of the Jerman function for all voxels, which is the most computationally expensive operation (due to eigenvalues' extractions), depends only the volume size. Consequently, the computational cost of the tracking part does not depend on the number of MB to pair, unlike Hungarian or graph-based algorithms, which is particularly well suited in high concentration environment.

Overall, the introduction of TAL as a novel approach addresses the challenges of tracking MB in high concentration environment and provides a physically motivated pairing strategy that considers the global continuity information available in a buffer. The method demonstrates superior performance in reconstructing hemodynamics, exhibiting robustness to noise variations, and enabling accurate velocity estimations.

These findings, supported by *in silico* and *in vivo* experiments, contribute to the advancement of subwavelength hemodynamics imaging techniques.

## V. Conclusion

We demonstrated the feasibility of recovering ULM density maps and extracting blood velocities in a range of vessels using a novel approach that reverses the localization and tracking steps. The newly proposed Tracking and Localization workflow presents a promising solution to enhance the reliability and accuracy of anatomic and functional results obtained through ULM and DULM techniques.